\documentstyle[epsfig,emulateapj]{article}

\def \ref {\noindent\hangindent=0.3in\hangafter=1}

\def\ltsima{$\; \buildrel < \over \sim \;$}
\def\simlt{\lower.5ex\hbox{\ltsima}} 
\def\gtsima{$\; \buildrel > \over \sim \;$}
\def\simgt{\lower.5ex\hbox{\gtsima}} 
\def\gsim{\mathrel{\hbox{\rlap{\lower.55ex \hbox {$\sim$}}
                   \kern-.3em \raise.4ex \hbox{$>$}}}}

\begin{document}

\title{BeppoSAX Observations of GRB980425: Detection of the Prompt Event
and Monitoring of the Error Box}

\author{
E. Pian\altaffilmark{1},
L. Amati\altaffilmark{1},
L. A. Antonelli\altaffilmark{2},
R. C. Butler\altaffilmark{1},
E. Costa\altaffilmark{3},
G. Cusumano\altaffilmark{4},
J. Danziger\altaffilmark{5},
M. Feroci\altaffilmark{3},
F. Fiore\altaffilmark{6},
F. Frontera\altaffilmark{1,7},
P. Giommi\altaffilmark{6},
N. Masetti\altaffilmark{1},
J. M. Muller\altaffilmark{6},
L. Nicastro\altaffilmark{4},
T. Oosterbroek\altaffilmark{8},
M. Orlandini\altaffilmark{1},
A. Owens\altaffilmark{8},
E. Palazzi\altaffilmark{1},
A. Parmar\altaffilmark{8},
L. Piro\altaffilmark{3},
J. J. M. in 't Zand\altaffilmark{9},
A. Castro-Tirado\altaffilmark{10},
A. Coletta\altaffilmark{6},
D. Dal Fiume\altaffilmark{1},
S. Del Sordo\altaffilmark{4},
J. Heise\altaffilmark{9},
P. Soffitta\altaffilmark{3},
V. Torroni\altaffilmark{6}
}

\altaffiltext{1}{Istituto Te.S.R.E., CNR, via Gobetti 101, I-40129
Bologna, Italy}

\altaffiltext{2}{Osservatorio Astronomico di Roma, sede di Monteporzio
Catone, Via Frascati 33, I-00040 Monteporzio Catone, Italy}

\altaffiltext{3}{I.A.S., C.N.R., Via Fosso del Cavaliere, Area della
Ricerca di Tor Vergata, I-00131 Rome, Italy}

\altaffiltext{4}{I.F.C.A.I., CNR, via Ugo La Malfa 153, I-90146 Palermo,
Italy}

\altaffiltext{5}{Osservatorio Astronomico di Trieste, Via G.B.
Tiepolo 11, I-34131 Trieste, Italy}

\altaffiltext{6}{BeppoSAX Scientific Data Center, Via Corcolle 19,
I-00131 Rome, Italy}

\altaffiltext{7}{Physics Department, University of Ferrara, Via Paradiso, 
     12, I-44100 Ferrara, Italy}

\altaffiltext{8}{Astrophysics Division, SSD of ESA, ESTEC, P.O. Box
299, 2200 AG Noordwijk, The Netherlands}

\altaffiltext{9}{Space Research Organization Netherlands,
Sorbonnelaan 2, 3584 CA Utrecht, The Netherlands}

\altaffiltext{10}{IAA-CSIC, Granada, Spain and LAEFF-INTA, Madrid,
Spain}


\begin{abstract}

We present BeppoSAX follow-up observations of GRB980425 obtained with
the Narrow Field Instruments (NFI)  in April, May, and November 1998. 
The first NFI observation has detected within the $8^{\prime}$ radius
error box of the GRB an X-ray source positionally consistent with the
supernova 1998bw, which exploded within a day of GRB980425, and a
fainter X-ray source, not consistent with the position of the supernova. 
The former source is detected in the following NFI pointings and
exhibits a decline of a factor of two in six months. If it is associated
with SN~1998bw, this is the first detection of X-ray emission from a
Type I supernova above 2 keV.  The latter source exhibits only
marginally significant variability.  The X-ray spectra and variability
of the supernova are compared with thermal and non-thermal models of
supernova high energy emission.  Based on the BeppoSAX data, it is not
possible to firmly establish which of the two detected sources is the
GRB X-ray counterpart, although probability considerations favor the
supernova. 

\end{abstract}

\keywords{gamma rays: bursts --- Supernovae: individual: SN~1998bw}

\section{Introduction}

The GRB of 1998 April 25, detected both by the BeppoSAX GRBM and by BATSE
(Kippen 1998) and localized with arcminute accuracy by the BeppoSAX
WFC (Soffitta et al. 1998), has received particular attention from the
astronomers because of its spatial (within a few arcminutes) and temporal
(within one day)  consistency with the optically and radio exceedingly
bright Type Ic supernova 1998bw (Galama et al.  1998;  Kulkarni et al.
1998a; Iwamoto et al. 1998), in the nearby galaxy ESO~184-G82 ($z =
0.0085$, Tinney et al. 1998).  The low probability of a chance coincidence
between the two events ($\sim 10^{-4}$, Galama et al. 1998) has
strengthened the hypothesis of a physical association between GRB980425
and the supernova.

However, since the other GRBs for which a redshift measurement is
available are located at larger distances ($z \sim 0.7$ or higher) and are
characterized by power-law decaying optical afterglows (Fruchter et al.
1999a; Djorgovski et al. 1997; Fruchter et al.  1999b; Halpern et al. 
1998;  Kulkarni et al. 1998b; Bloom et al. 1998a;  Vreeswijk et al. 1999; 
Castro-Tirado et al. 1999; Galama et al. 1999a;  Kulkarni et al. 1999; 
Fruchter et al. 1999c; Harrison et al. 1999;  Stanek et al.  1999; Bakos
et al. 1999), in agreement with the ``classical" fireball model (e.g.,
Rees \& M\'esz\'aros 1992; Piran 1999), GRB980425 has been regarded as a
possible representative of a separate GRB class, with apparently
indistinguishable high energy characteristics, but with different
progenitors. 

The existence of a particular class of GRBs and its possible association
with supernovae have been systematically searched for using BATSE catalogs
and supernovae compilations by several authors (Wang \& Wheeler 1998;
Norris et al. 1998; Bloom et al. 1998b; Kippen et al. 1998) or based on
individual cases of supernovae with outstanding optical properties
(Germany et al. 1999; Turatto et al. 1999; Terlevich, Fabian, \& Turatto 1999).

Following the detection of GRB980425, observations of its WFC error box
with the BeppoSAX Narrow Field Instruments (NFI) were immediately
activated, starting 10 hours and one week after the event.  The detection
of two previously unknown X-ray sources - one of which being consistent
with the position of the supernova, and the other possibly, but not
clearly, fading - was regarded as quite anomalous, because previous
BeppoSAX NFI follow-up observations of well localized GRBs had generally
detected X-ray transients characterized by power-law decay.  This fact
prompted further observations six months after the event, aimed at
clarifying the uncertainty about the GRB X-ray counterpart, and, as a
secondary though not less important scope, at monitoring the X-ray
emission of SN~1998bw, considering the peculiarity of this object and the
poor knowledge of the X-ray behavior in supernovae in general (see review
by Schlegel 1995), and particularly of Type I. 

In this Letter we present the high energy characteristics of the prompt
event as measured by the BeppoSAX GRBM and WFC, and the results of the
follow-up NFI observations (\S 2) and discuss their implications in view
of the detection of SN~1998bw in the GRB field (\S 3). A preliminary
report on these data has been given in Pian et al. (1999).

\section{Data analysis and results}

\subsection{Prompt event}

GRB980425 triggered the BeppoSAX GRBM at 21:49:11 UT, and was
simultaneously detected by the BeppoSAX WFC unit 2 (Soffitta et al. 
1998).  The event had a duration of 31 s in the range 40-700 keV and
40 s in the range 2-26 keV. It exhibited a single, non structured
peak profile in both bands, with the indication of a $\sim$5 s soft
lag (Fig.~1).  Some flux brightening and successive decrease, lasting
altogether $\sim$10 seconds, are 
seen in the WFC light curve after the first 40 seconds, but not
in the $\gamma$-rays. 

The GRBM and WFC light curves appear well correlated, with the
indication of a $\sim$5 seconds lag of the lower energies with respect
to the higher energies.  Figure 2 shows the Discrete Correlation
Function (DCF) between the two light curves.  This correlation method is
suited to search for correlations and temporal lags between two
discrete, and possibly unevenly sampled, data trains (Edelson \& Krolik
1988).  To a maximum of the DCF amplitude at a positive temporal lag
corresponds a positive correlation, with the higher energies leading the
lower ones.  A maximum at a positive temporal lag of $\sim$5 seconds is
evident.  The asymmetric shape of the DCF
function amplitude around its peak reflects the fact that the flux decay
after maximum is slower at X-rays than at $\gamma$-rays (see Fig. 1).

The fluences are ($2.8 \pm 0.5$)  $\times 10^{-6}$ erg cm$^{-2}$ and
($1.8 \pm 0.3$)  $\times 10^{-6}$ erg cm$^{-2}$ in the 40-700 keV and
2-26 keV energy range, respectively. (The Galactic absorption in the
direction of GRB980425, $N_{HI} = 4 \times 10^{20}$ cm$^{-2}$, from
Schlegel, Finkbeiner, \& Davis 1998, is negligible at energies higher than
2 keV.) The
spectral index of the $\gamma$-ray spectrum, averaged over the burst
duration, is $\alpha = 1.2 \pm 0.2$ ($f_\nu \propto \nu^{-\alpha}$), and
that of the X-ray spectrum is $\alpha = 0.41 \pm 0.25$.  Strong spectral
softening is evident during the event (Frontera et al.  1999). No
peculiar temporal or spectral characteristics are noted in this GRB.

The spacecraft aspect reconstruction conditions allowed us to only
poorly constrain the WFC error box of the GRB, which has a radius of
$8^{\prime}$.  This made the search of an X-ray transient more
difficult.  The IPN allowed a substantial reduction of the GRB error box
(see Galama et al. 1999b).

\subsection{NFI Target of Opportunity Observations}

The BeppoSAX NFI were pointed at the $8^{\prime}$ radius error box
determined by the WFC at three epochs starting 10 hours after the GRB:
April 26-28, May 2-3 and November 10-12 (see Journal of Observations in
Table~1; note that the April pointing was uninterrupted, but has been
split in two parts only for the data analysis purpose).  The strategy of
performing NFI pointings at so large time intervals after the GRB (one
week and six months), besides promptly thereafter, is not usually
adopted for GRBs and was dictated by the ambiguity of the results
obtained with the April observation, which suggested a quite different
case than previously observed X-ray afterglows. 

Event files for the LECS and MECS experiments were linearized and
cleaned with SAXDAS at the BeppoSAX Science Data Center (SDC; Giommi \&
Fiore 1998).  LECS (0.1-4 keV) and MECS (unit 2 and 3, 1.6-10 keV)
images for each pointing were extracted using the XIMAGE software
package.  The MECS images, better exposed and of higher signal-to-noise
ratio than the LECS ones, are reported in Fig.~3.

The analysis of the MECS imaging data of the first portion of the first
pointing (Fig.~3a) shows that inside the intersection of the WFC error
circle and IPN annulus, two point-like, previously unknown X-ray sources
are detected with a positional uncertainty of $1^{\prime}.5$:
1SAXJ1935.0-5248 (hereafter S1), at RA = 19h 35m 05.9s and Dec =
--52$^{\circ}$ 50$^{\prime}$ 03$^{\prime\prime}$, and 1SAXJ1935.3-5252
(hereafter S2), at RA = 19h 35m 22.9s and Dec = --52$^{\circ}$
53$^{\prime}$ 49$^{\prime\prime}$.  Note that the coordinates
distributed by Pian et al. (1998) have been revised in November 1998, to
take into account a systematic error due to the non-optimal spacecraft
attitude during the April and May 1998 observations (see Piro et al.
1998a). The revised position of S1 is consistent within the uncertainty
with the position of SN~1998bw detected in the WFC error box (Galama et
al.  1998;  Kulkarni et al.  1998a) and exploded simultaneously with the
GRB with an uncertainty of $\sim$1 day (Iwamoto et al.  1998), while the
revised position of S2 is $\sim 4^{\prime}.5$ away from SN~1998bw, and
therefore inconsistent with it (see Fig. 1 in Galama et al. 1999b). 

The LECS and MECS count rates and upper limits for both sources during the
three pointings have been computed within circles of 3$^{\prime}$ radius
and corrected for the local background estimated within circles of similar
size (Table~1). The upper limits have been estimated as explained in the
Appendix.  This method takes into account, in addition to the normal
photon statistics, also the fact that, at these flux levels, the LECS and
MECS background may be dominated by the fluctuations of the cosmic X-ray
background. 

Source S1 is detected by the MECS also in the following pointings at a
position consistent with that of the first observation.  The
observation of November 1998 shows a decrease in the X-ray flux of
approximately a factor of two with respect to the level measured in
April-May 1998.  The source is also detected by the LECS in the April
pointing, and not in the November pointing (LECS data of the May 1998
pointing are excluded from the analysis because of the extremely short
exposure time).

Due to the limited spatial resolution of the LECS and MECS detectors and
to the faint emission level of source S2, estimating its flux is made
difficult by the background contamination and by the proximity of the
brighter source S1. During the second portion of the April pointing, as
well as in the November 1998 pointing, S2 is not detected by the MECS,
while signal from a position consistent with that of S2 in April is
marginally detected in the May 1998 pointing (Fig.~3c), with 
lower flux than in the first observation (see Table~1), but consistent
with it at the $\sim$2-$\sigma$ level. We note that the May detection,
albeit of a signal-to-noise formally lower than 3, has a very low
probability of being a background fluctuation ($\sim 10^{-6}$) when
considered together with the April detection.  However, we
conservatively report in Table 1 also the 3-$\sigma$ upper limit for the
May
measurement. All upper limits to the flux of S2 are consistent with the
level of the detection. There is no significant detection of S2 in the
individual LECS images. 

No significant signal above background is detected by the BeppoSAX high
energy instruments PDS and HPGSPC in any of the three pointings. 
 
Light curves and spectra for each pointing were accumulated with the
XSELECT tool, using a 3 arcmin extraction radius both for the LECS and
the MECS, which provides only 80\% of flux, but allows partially avoiding
the mutual contamination of S1 and S2.  Since the local background
intensity is similar to that measured from files accumulated from blank
fields available at the SDC, we used the latter, which are affected by a
smaller uncertainty.  No significant variability within each pointing is
exhibited by either source.

Spectral analysis of the LECS and MECS data has been done with the XSPEC
10.0 package using the response matrices and auxiliary files available at
the SDC. LECS and MECS spectra flux distributions of both sources have
been grouped in order to achieve a signal-to-noise ratio of at least
$\sim$3 in each bin.

For S1, a fit with a single power-law ($F_\nu \propto \nu^{-\alpha}$) 
absorbed by Galactic extinction of the individual MECS spectra is
satisfactory, with reduced $\chi^2$ values well below one, due to the
large errors.  For these fits at individual epochs the LECS spectra have
not been used because they have too low signal-to-noise ratio. 

Since no significant spectral variability is seen from epoch to epoch,
we averaged the LECS and MECS spectra of April and May, to increase the
signal-to-noise ratio, and fitted a single power-law to the average
rebinned spectrum. We excluded the November spectrum from this average
because the count rates indicate that the flux varied at that epoch
with respect to April-May (Table 1). 

We obtain a spectral index of $\alpha = 1.0 \pm 0.3$ (all fit parameters for
the average spectrum are reported in Table~2). The fitted $N_{HI}$ is found to
be consistent with the Galactic value and therefore we have fixed it to that
value.  The intercalibration constant between the LECS and MECS data is within
the expected range.  However, the fit is formally not completely satisfactory
(reduced $\chi^2$ = 1.2, see Table 2), due to a flux excess at energies below
$\sim 1$ keV (Fig. 4a). 

We also tried a fit with a thermal bremsstrahlung with Galactic
absorption, obtaining a temperature $kT \simeq 8$ keV and a not
satisfactory reduced $\chi^2$ of 1.7, still due to the presence of a
soft excess (Fig 5a). 

Fitting the data assuming no Galactic absorption still yields some soft
excess with respect to both a power-law and a bremsstrahlung model,
therefore we tend to exclude that the effect is due to an overestimate
of the Galactic extinction in the direction of SN~1998bw (as possibly
suggested by Patat et al. 1999).

A fit of the data with a broken power-law and Galactic absorption 
yields an index $\alpha_1 = 2.0 \pm 0.5$ below 1.4 keV and 
$\alpha_2 = 0.7 \pm 0.4$ at the higher energies.  The reduced $\chi^2$
is 0.5, considerably lower than for the single power-law and for the
bremsstrahlung models (see Fig. 6).

To account for the soft excess we also tried composite fits of a
power-law or a thermal bremsstrahlung with a black body model plus
Galactic
absorption.  The former fit yields a spectral index $\alpha = 0.5 \pm
0.3$ and a black body temperature $kT = 90 \pm 20$ eV, with a reduced
$\chi^2 = 0.5$ (Fig. 4b). The latter yields a bremsstrahlung temperature
of $kT \sim 16$ keV (see Table 2), and black body temperature and
normalization consistent with those found in the former case (reduced
$\chi^2 = 0.5$, Fig.  5b). 

We found no obvious reason for the excess to be spurious, such as
instrumental effect or non optimal background subtraction.  Therefore we
considered it real, although not highly significant (see Figures 4a and
5a). The chance probabilities that adding a black body component either to
a power-law or to a thermal bremsstrahlung model improves the fit are
1.3\% and 0.17\%, respectively. Fitting the data with a broken instead of
a single power-law has a chance probability of improvement of 4\%.

The low signal-to-noise ratio of the LECS data in November 1998 does not
allow us to test the goodness of the fit of those data with a black
body model.  By fixing the black body temperature to the best fit value
of the April-May spectrum, 90 eV, and fitting a power-law plus black
body model to the November MECS spectrum and LECS upper limit we get a
black body normalization upper limit consistent with the value obtained
for April-May, indicating that this component has not varied
significantly.  The power-law normalization has instead varied by a
factor of $\sim 2$. 

Although there is no detection of S2 in the individual LECS images,
signal is present in the April-May coadded image.  Therefore, since the
MECS spectra of the individual observations have a very low
signal-to-noise ratio, we only used the rebinned LECS+MECS spectrum
obtained from the average April-May image, and fitted it with a power-law
of index $\alpha = 1.5 \pm 0.4$ plus Galactic absorption ($\chi^2 = 0.5$). 

Due to the high uncertainty in the power-law spectral index and to the low
level of both S1 and S2, we estimated their intensities in the 2-10 keV
band by assuming a standard factor of conversion from count rates of $9.3
\times 10^{-11}$, corresponding to an adopted power-law spectral shape of
index 0.5.  The 1-$\sigma$ uncertainties on the intensities have been
obtained by similarly scaling the errors on the count rates (Table 1). 
These intensities are reported in Figure 7, along with the WFC light curve
in the 2-10 keV band, obtained by binning in intervals of 5-20 seconds the
temporal profile reported in Figure 1.  Differences between these
intensities and those obtained by adopting a bremsstrahlung model do not
exceed 10\%. Note that the contribution of the black body component to the
intensity of S1 in the 2-10 keV range is negligible.

\section{Discussion} 

\subsection{Light curve and spectral shape of source S1}

The X-ray light curve of source S1 measured by the NFIs shows a decay of
a factor of two in $\sim$6 months (Fig.~7b), much slower than X-ray GRB
afterglows so far observed. Assuming, as suggested by the positional
coincidence and by variability, that S1 is associated with SN~1998bw,
this is the first detection of medium energy X-ray emission from a Type
I supernova (there is a unique case of Type I supernova detected in soft
X-rays, the Type Ic SN~1994I, Immler et al.  1998a) and the earliest
detection of X-rays after supernova explosion. 

At the distance of SN~1998bw, 38 Mpc, the luminosity observed in the
range 2-10 keV, $\sim 4-7 \times 10^{40}$ erg s$^{-1}$, would be
compatible with that of other supernovae detected in the same energy band
(Kohmura et al. 1994; Houck et al. 1998; Schlegel 1995, and references
therein).  
 
However, the observed luminosity and variation thereof represent only
an upper limit to the luminosity and a lower limit to the amplitude of
X-ray variability of SN~1998bw, respectively, due to the possible
contribution of its host galaxy, a face-on spiral galaxy about one
tenth of the size of our Galaxy, which is only very marginally
resolved in the BeppoSAX data. In fact, a galaxy of that type and size
could easily account for almost all of the X-ray emission observed in
November 1998, when the flux was lowest (see e.g., Fabbiano 1989). 

The observed decay of S1 in the 2-10 keV band is well fitted by a
power-law $F(t) \propto t^{-p}$ with $p = 0.16 \pm 0.04$ (reduced
$\chi^2 \simeq 0.7$).  The fit with an exponential law $F(t) \propto
e^{-t/\beta}$ with $\beta = 500 \pm 100$ days has a reduced $\chi^2$ of
2.4, corresponding to a probability of $\sim$10\% for 2 degrees of
freedom, therefore not negligible (Fig.~7b).  Both trends would be
similar, considering the unknown dilution by the host galaxy of
SN~1998bw, to the X-ray behavior of other supernovae (e.g., Kohmura et
al.  1994;  Zimmermann et al. 1994; Houck et al. 1998) and predicted by
models of thermal bremsstrahlung of energetic electrons within the
circumstellar medium (see e.g., Chevalier \& Fransson 1994; Chugai \&
Danziger 1994).

The prompt X-ray emission observed for SN~1998bw requires that the
circumstellar medium is highly ionized (perhaps by the powerful
explosion), to allow the X-rays to escape so soon after the explosion (see
Zimmermann et al. 1994), and also very dense, as inferred also from the
large radio output (Kulkarni et al. 1998a;  Wieringa, Kulkarni, \& Frail 
1999). In
these conditions, the X-rays might be produced by the reverse shock
which results from the pressure of swept-up material on the outgoing shock
and propagates back into the shocked supernova gas (Chevalier \& Fransson
1994; Schlegel 1995). 

The temperature obtained from the thermal bremsstrahlung fit to the
BeppoSAX LECS+MECS spectra, admittedly very poorly constrained, is
compatible with that fitted to the medium or hard X-ray spectra of other
supernovae (Kohmura et al. 1994; Leising et al. 1994; Dotani et al. 
1987). 

On the other hand, the mildly relativistic conditions evidently present
in the expanding shock of SN~1998bw at early epochs (Kulkarni et al.
1998a) and the
acceptable power-law fit obtained for the medium energy X-ray spectra
might suggest that non-thermal mechanisms are responsible for the X-ray
emission such as synchrotron radiation by the extremely energetic
electrons, as was modeled for the radio emission by Li and Chevalier
(1999), or inverse Compton scattering of relativistic electrons off
optical/UV photons of the thermal ejecta (see Canizares et al. 1982). 
The X-ray spectral index is consistent with that measured for the radio
spectrum starting $\sim 15$ days after the explosion (Kulkarni et al.
1998a;  Wieringa et al. 1999), and with the slope connecting
quasi-simultaneous radio and X-ray measurements ($\alpha \sim 0.8$). 
Therefore, it is difficult to establish whether the X-rays are produced
through synchrotron or inverse Compton radiation.  However, if inverse
Compton losses were dominant, radio emission production would be rapidly
inhibited (Schlegel 1995), contrary to the observations. 

Assuming the X-rays have a synchrotron origin and adopting a single
power-law of index $\alpha \sim 0.8$ for the radio-to-X-ray synchrotron
spectrum, we obtain a bolometric luminosity of $\sim 9 \times 10^{40}$ erg
s$^{-1}$. Assuming that the radio and X-ray emitting regions are cospatial
and expanding with a speed of $\sim 0.3c$ (Kulkarni et al.  1998a), we
derive a magnetic field of $\simlt 1$ Gauss, similar to that found for
SN~1980K by Canizares et al. (1982) assuming inverse Compton losses are
responsible for the X-ray production. 

However, the limited signal-to-noise ratio of the BeppoSAX spectra does
not allow us to choose between synchrotron radiation and thermal
bremsstrahlung as the mechanism for the medium energy X-ray production.

The emission component detected in the softer part of the BeppoSAX
spectrum, if fitted with a black body, has a temperature of $\sim$0.1
keV, corresponding to a black body total luminosity of $\sim 10^{41}$ erg
s$^{-1}$. The inferred linear size of the emitting region is about one
third of the solar radius, too large for a compact object left as a
remnant of the supernova explosion, but approximately compatible with the
size of the putative accretion disk promptly formed as a consequence of
the ``hypernova" or ``collapsar" phenomenon, of which SN~1998bw might be
an example (Paczy\'nski 1998;  MacFadyen \& Woosley 1998;  Woosley,
Eastman, \& Schmidt 1999a). However, such a compact object or disk could
be hardly
visible at a so early epoch, due to the optical thickness of the material
enshrouding it.

Black body soft X-ray emission is not expected from the supernova itself
or from the expanding shell or ejecta, due to non-equilibrium conditions
of the system.  However, the occurrence of short ($\sim$1000 s) thermal
bursts at early times after supernova explosion has been predicted,
albeit at very soft energies, possibly even lower than those observable by
the LECS (Schlegel 1995). We found no evidence in the LECS light curves
of similar events, but the sampling is not conducive to that detection. 

However, since there is no evidence of variability of the soft component
in the long term, it might not be related to the supernova. 
In fact, its spectrum can be fitted also with a power-law superimposed
on, and steeper than, the one which describes the spectrum at higher
energies. This suggests that the component might have a more complex
spectrum (possibly extending toward ultraviolet wavelengths), of which a
black body or steep power-law are only approximations. 
It could rather be a persistent (or slowly variable
with a small amplitude) source of soft X-rays such as the host galaxy
itself, or just its bulge, or the superposition of unresolved X-ray
sources within that galaxy, or diffuse hot gas, or the HII region in
which SN~1998bw is located (Galama et al. 1998), or the underlying
cluster DN 1931-529, or more probably, the sum of some or all of these
contributions.  The limited angular resolution of BeppoSAX does not
allow us to disentangle this component from the supernova itself.
Notwithstanding this possibility, we note that the unabsorbed luminosity
in the 0.1-2 keV range, $5 \times 10^{40}$ erg s$^{-1}$, given by the
superposition of the fitted power-law and black body components, is
similar to luminosities of supernovae observed in soft X-rays (Canizares
et al.  1982; Bregman \& Pildis 1992;  Zimmermann et al.  1994; 
Schlegel et al.  1996;  Fabian \& Terlevich 1996;  Immler et al. 
1998b).

\subsection{The association between GRB980425 and SN~1998bw}

The GRB980425 prompt event is relatively weak with respect to other
GRBs, and rather soft. However, it has no outstanding features with
respect to other BeppoSAX or BATSE GRBs, which might suggest a peculiar
counterpart at longer wavelengths, such as a bright supernova, instead
of a ``classical"  power-law fading afterglow.  The $\sim$5-seconds
temporal lag between the WFC and the GRBM light curves could be due to a
delay of X-ray emission during the burst with respect to the
$\gamma$-rays, or ascribed to intrinsic absorption in a medium becoming
increasingly transparent (see e.g., B\"ottcher et al.  1999).  Similar
soft lags from few to $\sim$10 seconds are observed also in other GRBs
(Piro et al. 1998b; Piro et al. 1998c; Frontera et al.  1999). 

If SN~1998bw is the counterpart of GRB980425, the production of
$\gamma$-rays could be accounted for by the explosion of the 14
$M_{\odot}$ helium core of a $\sim 35 ~ M_{\odot}$ star (Woosley et al.
1999a; MacFadyen \& Woosley 1999)  and by the subsequent expansion of a
relativistic shock, in which non-thermal electrons are radiating photons
of $\sim$100 keV, provided the explosion is asymmetric, i.e. the GRB is
produced in a relativistic jet (Iwamoto et al.  1998;  Woosley et al. 
1999a; H\"oflich, Wheeler, \& Wang 1999; Rej 1998; see however, Kulkarni
et al.
1998a). The presence of an undetectable, or barely detectable,
non-thermal GRB remnant, underlying the brighter thermal supernova
ejecta cannot be excluded (see e.g., Iwamoto 1999).

Recent speculations have led to the proposal that every long ($> 1$ s) GRB
is formed via supernova, or hypernova, explosion (MacFadyen \& Woosley
1999).  The presence of a supernova underlying the GRB afterglow has been
recently tested for the optical transients of some GRBs with suggestive
results (GRB970228, Reichart 1999; Galama et al. 1999c;  GRB970508,
Germany et al. 1999;  GRB980326, Bloom et al. 1999; GRB990510, Fruchter et
al.  1999d; Beuermann et al.  1999; GRB990712, Hjorth et al.  1999). 
Indeed, the recent discovery of a GRB optical counterpart at the
intermediate redshift $z = 0.43$ (Galama et al. 1999d) might support a
continuity of properties between GRB980425 and the other precisely
localized GRBs, perhaps based on the different amount of jet collimation
(Woosley, MacFadyen, \& Heger 1999b) or different beaming, depending on
the degree of jet alignment (Eichler \& Levinson 1999; Cen 1998;
Postnov, Prokhorov, \& Lipunov 1999). In highly collimated or highly
beamed GRBs the non-thermal multiwavelength
afterglow could overwhelm the underlying supernova emission. This should
instead be detected more clearly in less collimated or less beamed
(i.e., seen
off-axis) GRBs, like GRB980425, which are, or appear, weaker.
Assuming
association
with SN~1998bw and isotropic emission, the total energy of GRB980425 in
the 40-700 keV, $\sim 5 \times 10^{47}$ erg, is at least four orders of
magnitude less than that of GRBs with known distance.

On the other hand, disregarding the fact that the probability of a
chance coincidence of GRB980425 and SN~1998bw is extremely low, one
might consider S2 as the X-ray counterpart candidate of the burst. 
The possible detection of S2 in 2-3 May 1998, one week after the GRB, 
implies, with respect to the first NFI detection in April, a much
slower decay than that normally observed for X-ray afterglows
(e.g., Costa et al. 1997;  Nicastro et al. 1998; Dal
Fiume et al. 1999; in 't Zand et al. 1998;  Nicastro et al. 1999;
Vreeswijk et al. 1999;  Heise et al. 1999). 

Assuming a power-law decay between the X-ray flux measured by the WFC in
the 2-10 keV range in the last $\sim$20 seconds of the GRB and the flux
measured in the first NFI observation (Fig. 7a), we derive a power-law
index $p \sim 1.5$, which is similar to commonly observed X-ray
afterglows.  If S2 is an afterglow, one would expect that its intraday
variability followed this same temporal behavior.  Therefore, we have
binned the light curve of S2 in the first portion of the April 1998
pointing in five intervals of 20000 seconds each.  We have then connected
with power-laws the last WFC measurement with the
first and last of these fluxes and have determined their indices to be $p
\simeq 1.6$ and $p \simeq 1.4$ (Fig. 7c), respectively.  The reduced
$\chi^2$ values computed for these two power-laws with respect to the
remainder four NFI data points of April are 3 and 30, respectively,
corresponding to low probabilities (1\% and $\ll$1) that the power-laws
describe the observed intrapointing light curve. (All points seem rather
consistent with a constant trend.)

The upper limit derived for the second portion of the April pointing is
inconclusive.  However, the detection of S2 in May 1998 suggests a
marginal deviation from the above power-laws ($\simgt 2.5 \sigma$). 
Therefore, the present data exclude at a confidence level of $\sim$99\%,
or higher, that S2 is an afterglow, unless a small re-bursting, one week
after the GRB, is superimposed
to the power-law monotonic decline.  This would be reminiscent of
GRB970508, although the time scales for re-bursting occurrence and
duration would be very different (Piro et al. 1998b).

\section{Conclusion}

Two previously unknown X-ray sources have been detected by the BeppoSAX
NFIs in the field of GRB980425. Neither of them has the obvious
characteristics of an X-ray afterglow, when compared to previously
observed cases.

SN~1998bw is a very interesting candidate for further monitoring in the
X-rays.  Thanks to its rapid slew capability and to its wide energy
range, BeppoSAX has promptly measured its spectrum up to energies beyond
$\sim 5$ keV, where supernovae have been so far largely unexplored.
Unfortunately, the signal is relatively modest (this is the second most
distant supernova detected so far in the X-rays, after SN~1988Z, located
at 95 Mpc, Fabian \& Terlevich 1996), and therefore longer exposures and
data of better signal-to-noise ratio are necessary to study in detail
this source and its environment.

Concerning the identification of the X-ray counterpart of GRB980425, our
tentative conclusion is that S1 has a high probability of being
associated with GRB980425, while S2 is more probably a variable field
source, albeit constant in the long term, like an active galaxy or a
Galactic X-ray binary (we note that the probability of detecting by chance
a source of the level of S2 is rather high, $\sim$10\%, Cagnoni et al.
1998; Giommi et al. 1998).  A spectroscopic survey of the NFI
error box of S2 has been inconclusive in this respect (Halpern 1998). 
Observations of this field by an X-ray instrument with higher
sensitivity (e.g., XMM) and spatial resolution (e.g., Chandra) than
those attained by BeppoSAX might help elucidating this controversial
issue.

\acknowledgements

We thank the BeppoSAX Mission Planning Team and the BeppoSAX SDC and SOC
personnel for help and support in the accomplishment of this project.

\newpage
\appendix
\section{Estimate of Upper Limits at 3-$\sigma$ confidence level}

Suppose that $m$ counts are found in the region where the target is
expected to appear and that in the same search area $b$ counts are
expected from the background. If $m - b < 3\sqrt b$ we have no positive
detection (at the 3-$\sigma$ confidence level) of the source and
a 3-$\sigma$ upper limit is needed.  We define the 3-$\sigma$ upper
limit as the number $x$ that gives a probability to observe $m$ or less
counts equal to the formal 3-$\sigma$ Gaussian probability, i.e.:

$$ P(\le m, x+b) = P_{Gauss} (3\sigma) \eqno(1)$$

Assuming Poisson statistics, eq. (1) becomes:


$$ e^{-(x+b)} \sum_{i=1}^{m} \frac{(x+b)}{i!} = 2.7 \times
10^{-3}
\eqno(2) $$

In the limit of large numbers eq. (2) reduces to

$$ m = x + b - 3 \sqrt{x+b} \eqno(3) $$

By solving (3) with respect to $x$  we have 


$$ x = \frac{9 + 2m + 3\sqrt{9 + 4m} - 2b}{2}. $$

\newpage


\newpage


\begin{center}
\begin{tabular}{rcccccc}
\multicolumn{7}{c}{{\bf Table 1:} Journal of BeppoSAX LECS and MECS
Observations}\\
\hline
\hline
\multicolumn{1}{c}{Date (UT)} & \multicolumn{3}{c}{LECS} &
\multicolumn{3}{c}{MECS} \\
& t$^a$ (s) & \multicolumn{2}{c}{Flux$^b$($\times 10^{-3}$ cts s$^{-1}$)} &
t$^a$ (s) & \multicolumn{2}{c}{Flux$^c$($\times 10^{-3}$ cts s$^{-1}$)}
\\
& & S1 & S2 & & S1 & S2 \\
\hline
1998 Apr 26.334-27.458 & 24483 & $4.0 \pm 1.3^d$ & $< 4.0$ &
                         37220 & $4.6 \pm 0.6$ & $2.4 \pm 0.5$ \\
     Apr 27.469-28.160 & 13566 & $< 7.0$ & $< 7.0$ &
                         21805 & $4.5 \pm 0.7$   & $< 2.5$ \\
     May 02.605-03.621 & 1016.5 & -- & -- &
                         31975 & $3.0 \pm 0.5$   & $1.4^e \pm 0.5$ \\
     Nov 10.754-12.004 & 16961 & $< 6.0$ & $< 6.0$ &
                         53122 & $1.8 \pm 0.4$   & $< 2.0$ \\
\hline
\multicolumn{7}{l}{$^a$ On source exposure time}\\ 
\multicolumn{7}{l}{$^b$ In the energy range 0.1-4 keV}\\
\multicolumn{7}{l}{$^c$ In the energy range 1.6-10 keV}\\
\multicolumn{7}{l}{$^d$ All uncertainties are at 1-$\sigma$; upper
limits are at 3-$\sigma$}\\
\multicolumn{7}{l}{$^e$ The 3-$\sigma$ upper limit is 
$1.9 \times 10^{-3}$ cts s$^{-1}$}\\
\end{tabular}
\end{center}

\vspace{2cm}

\begin{center}
\begin{tabular}{cccccccc}
\multicolumn{8}{c}{{\bf Table 2:} Fits to the BeppoSAX LECS+MECS
Average Spectrum of S1 in April-May 1998}\\
\hline  
\hline
Model & $\alpha_1$ & $\alpha_2$ & $E_{break}$ & $kT_{THB}$ & $kT_{BB}$
&
Red. $\chi^2$ & D.o.F.\\
& & & keV & keV & eV & & \\
\hline  
Single Power-Law & $1.0 \pm 0.3$ & - & - &- & - & 1.18 & 12 \\
Thermal Bremsstrahlung & - & - & - & $7.5^{+13.5}_{-3.5}$ & - & 1.66 & 12
\\
Broken Power-Law & $2.0 \pm 0.5$ & $0.7 \pm 0.4$ & $1.4 \pm 0.6$ & - &
- & 0.49 & 9 \\
Power-Law + Black Body & $0.5 \pm 0.3$ & - & - & - & $90^{+30}_{-20}$ &
0.49 & 10 \\
Th. Bremss. + Black Body & - & - & - & $16^{+130}_{-10}$ &
$90^{+30}_{-20}$ & 0.46 & 10 \\
\hline
\multicolumn{7}{l}{$^a$ Uncertainties are at 90\% (1.6-$\sigma$)
confidence level}\\
\end{tabular}
\end{center}


\newpage

\figcaption{BeppoSAX WFC (top) and GRBM (bottom) light curves of
GRB980425.  The onset of the GRB, indicated by the zero abscissa,
corresponds to 1998 April 25.909097 (i.e., 5 seconds earlier than the
GRBM trigger time).  The vertical bars represent the typical 1-$\sigma$
uncertainty associated with the individual flux points.}

\figcaption{Discrete Correlation Function between WFC and GRBM light
curves.  The maximum at $\sim$5 seconds indicates positive correlation
between the two curves with the WFC light curve lagging the GRBM one by
that temporal lag.}

\figcaption{BeppoSAX MECS images of the field of GRB980425 in the energy
band 1.6-10 keV referring to the epochs: (a) 1998 April 26.334-27.458; 
(b) April 27.469-28.160; (c) May 2.605-3.621; (d) November 10.754-12.004.
The local background has been subtracted and a smoothing has been applied
with a window function of width comparable to the MECS detector point
spread function (3$^{\prime}$ at half power diameter).  The source S1 is
clearly seen in all images. The fainter source S2 is detected in the first
part of the first pointing, panel (a), and in the May 1998 pointing, panel
(c).  Note that in May 1998, panel (c), the pointing center was
significantly displaced with respect to the April pointing, which accounts
for the presence of the bright source toward North-East, about
$9^{\prime}$ from the image center, not visible in panels (a) and (b).  In
each image, the most external contours represent 2-$\sigma$ flux levels.}

\figcaption{BeppoSAX LECS and MECS spectra of S1 fitted with (a) a single
power-law and (b) a power-law plus black body.  The lower panels show the
ratios between the data and the model. See Table 2 for the fit
parameters.}

\figcaption{BeppoSAX LECS and MECS spectra of S1 fitted with (a) a thermal
bremsstrahlung and (b) a thermal bremsstrahlung plus black body.  The
lower panels show the ratios between the data and the model. See Table 2
for the fit parameters.}

\figcaption{BeppoSAX LECS and MECS spectra of S1 fitted with a broken
power-law. The lower panel shows the ratios between the data and the
model. See Table 2 for the fit parameters.}

\figcaption{BeppoSAX MECS light curves in the 2-10 keV band of the X-ray
sources S1 (open squares) and S2 (filled circles) detected in the
GRB980425 field. The WFC early measurements in the same band are also
shown (stars).  Uncertainties for the NFI measurements are 1-$\sigma$. 
The 1-$\sigma$ error bars for the WFC points are smaller than the symbol
size and have not been reported. (b) Same as (a) for source S1 only. The
fits to the temporal decay with a power-law of index $\sim 0.2$ and with
an exponential law of $e$-folding time $\sim 500$ days are shown as dotted
and dashed curves, respectively. (c) Same as (a) for source S2 only. The
first NFI measurement of S2 in (a) is replaced here by 5 points obtained
by integrating and averaging the flux in shorter time intervals. The
dotted lines represent the power-laws of indices $p \simeq 1.6$ and $p
\simeq 1.4$ connecting the last WFC measurement and the first and last of
the 5 NFI points of the April 26-27 light curve, respectively.  The
extrapolations of the power-laws to the time of the third NFI observation
(May 1998) fall below the lower bound of the data point, although they are
compatible with it at the $\simgt 2.5 \sigma$ level.}

%

\newpage

\null
\vspace{-3cm} 

\begin{figure}

\epsfysize=12cm 
\hspace{3cm}\epsfbox{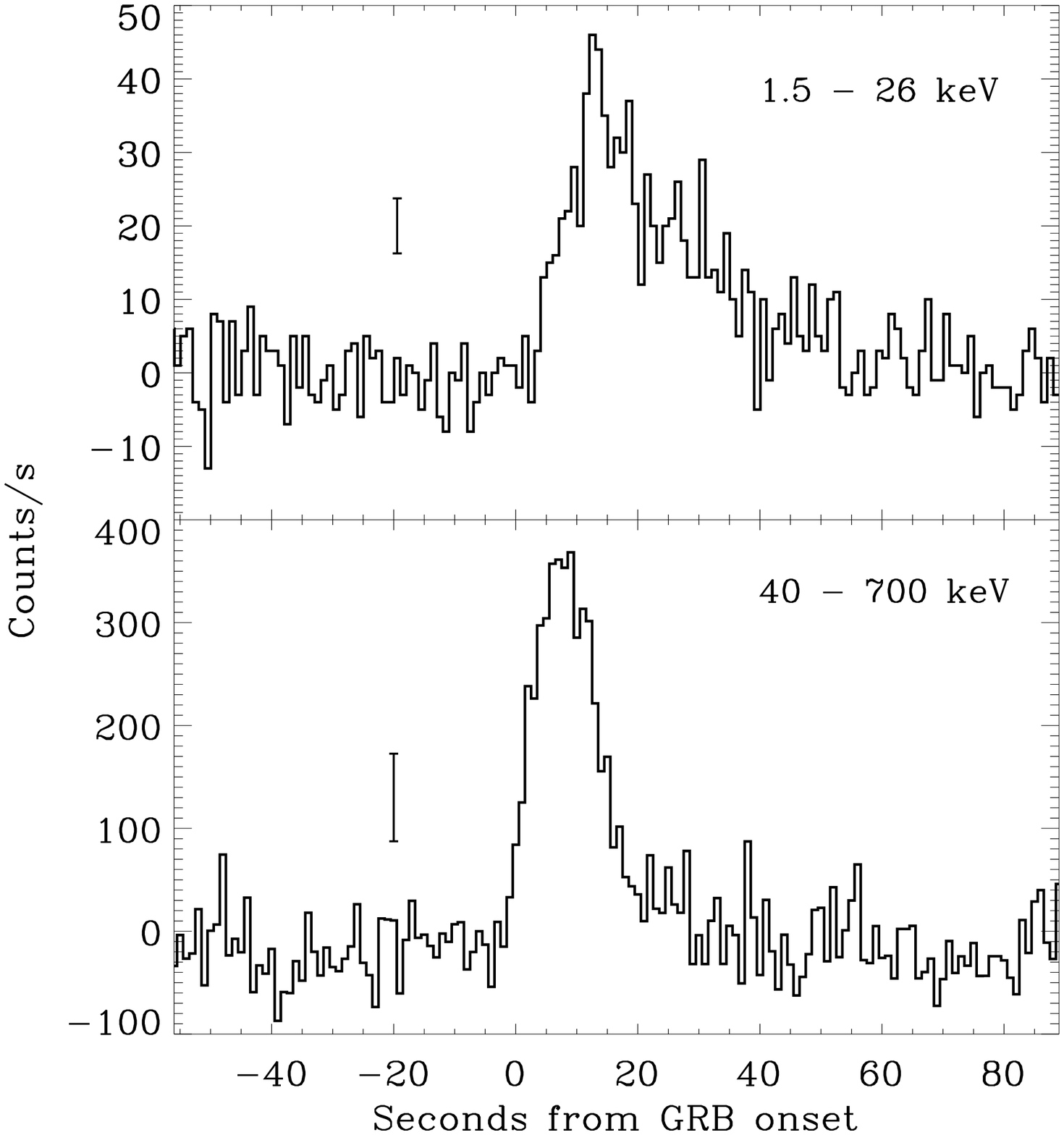} 

\vspace{2cm}

Fig. 1 -- BeppoSAX WFC (top) and GRBM (bottom) light curves of
GRB980425.  The onset of the GRB, indicated by the zero abscissa,
corresponds to 1998 April 25.909097 (i.e., 5 seconds earlier than the
GRBM trigger time).  The vertical bars represent the typical 1-$\sigma$
uncertainty associated with the individual flux points.

\end{figure}  


\newpage

\begin{figure}
\epsfysize=15cm 
\hspace{1cm}\epsfbox{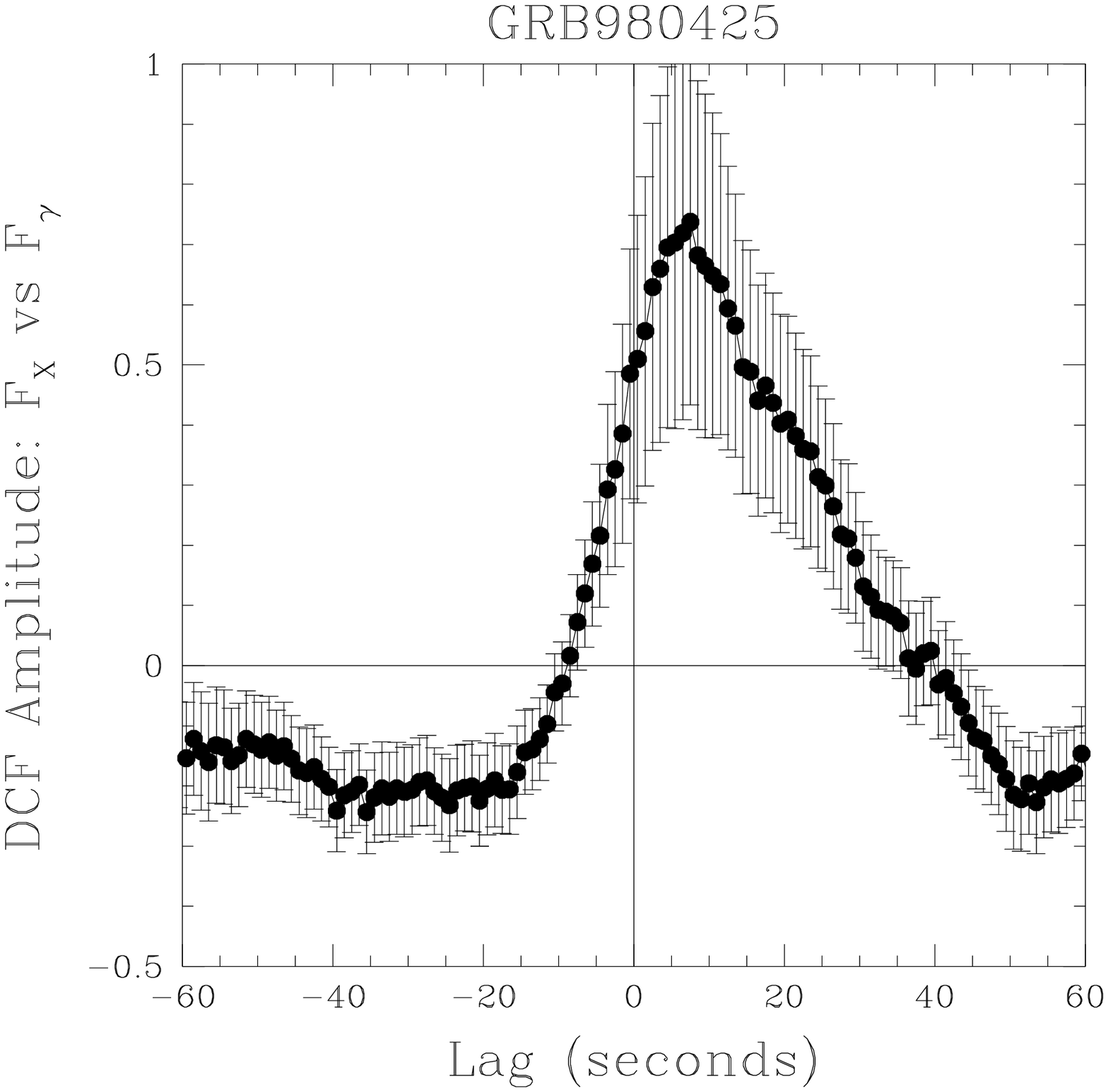} 

Fig. 2 -- Discrete Correlation Function between WFC and GRBM light
curves.  The maximum at $\sim$5 seconds indicates positive correlation
between the two curves with the WFC light curve lagging the GRBM one by
that temporal lag.

\end{figure}


\newpage
\null


\begin{figure}
\epsfxsize=9cm 
\hspace{3cm}\epsfbox{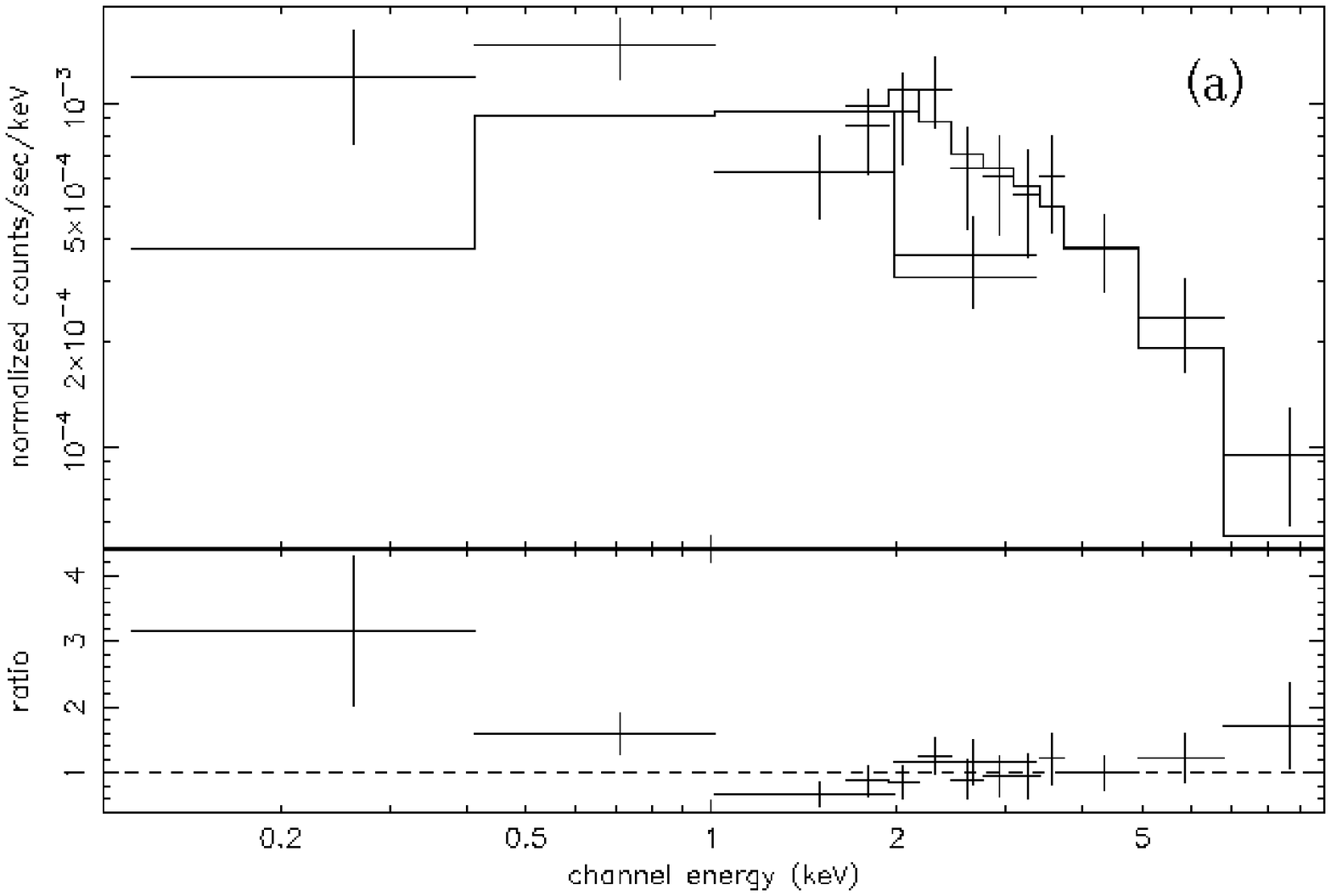} 

\vspace{2cm}

\epsfxsize=9cm 
\hspace{3cm}\epsfbox{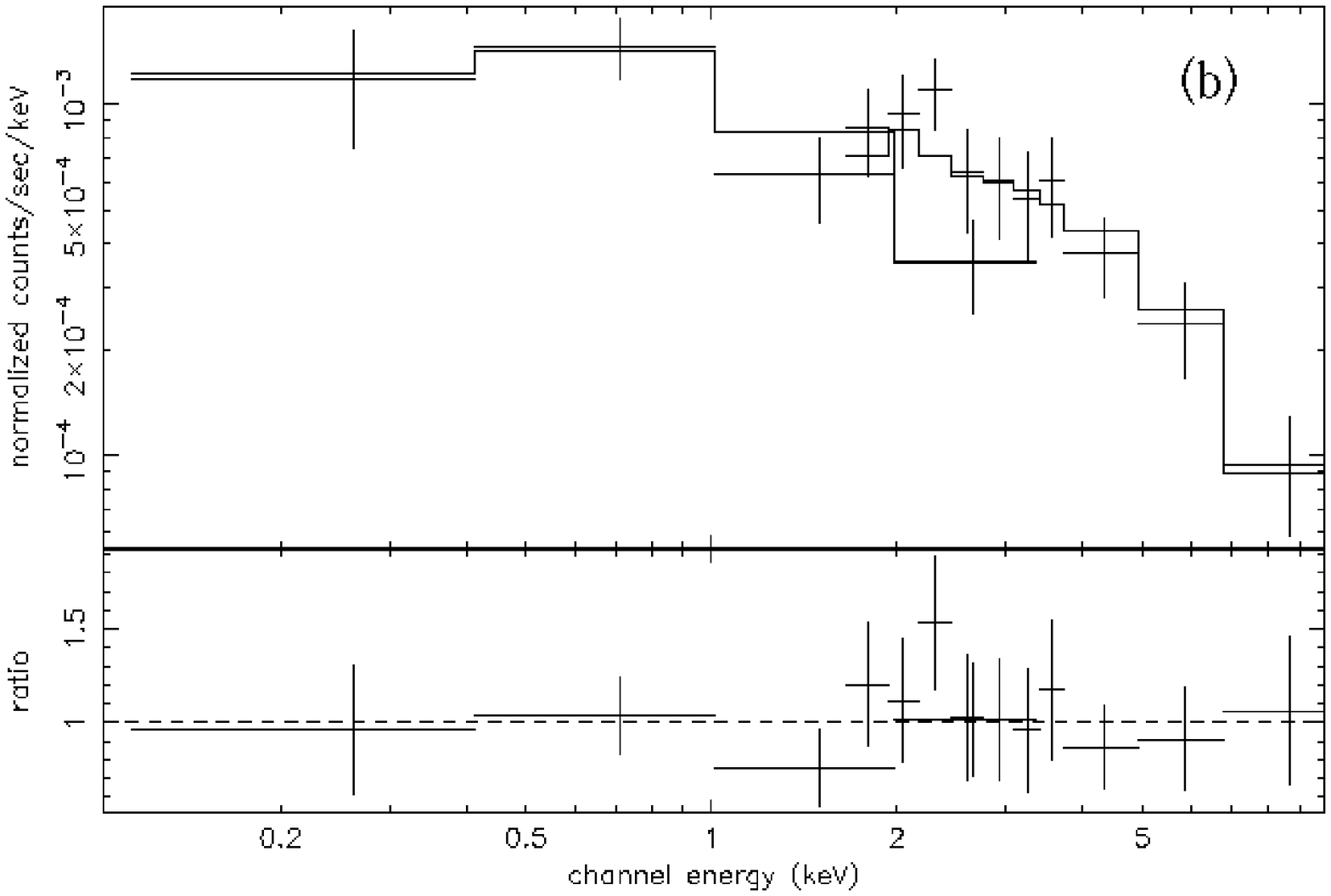} 

\vspace{3cm}
  
Fig. 4 -- BeppoSAX LECS and MECS spectra of S1 fitted with (a) a single
power-law and (b) a power-law plus black body.  The lower panels
show the ratios between the data and the model. See Table 2 for the
fit parameters.

\end{figure}

\newpage

\begin{figure}


\epsfxsize=9cm 
\hspace{3cm}\epsfbox{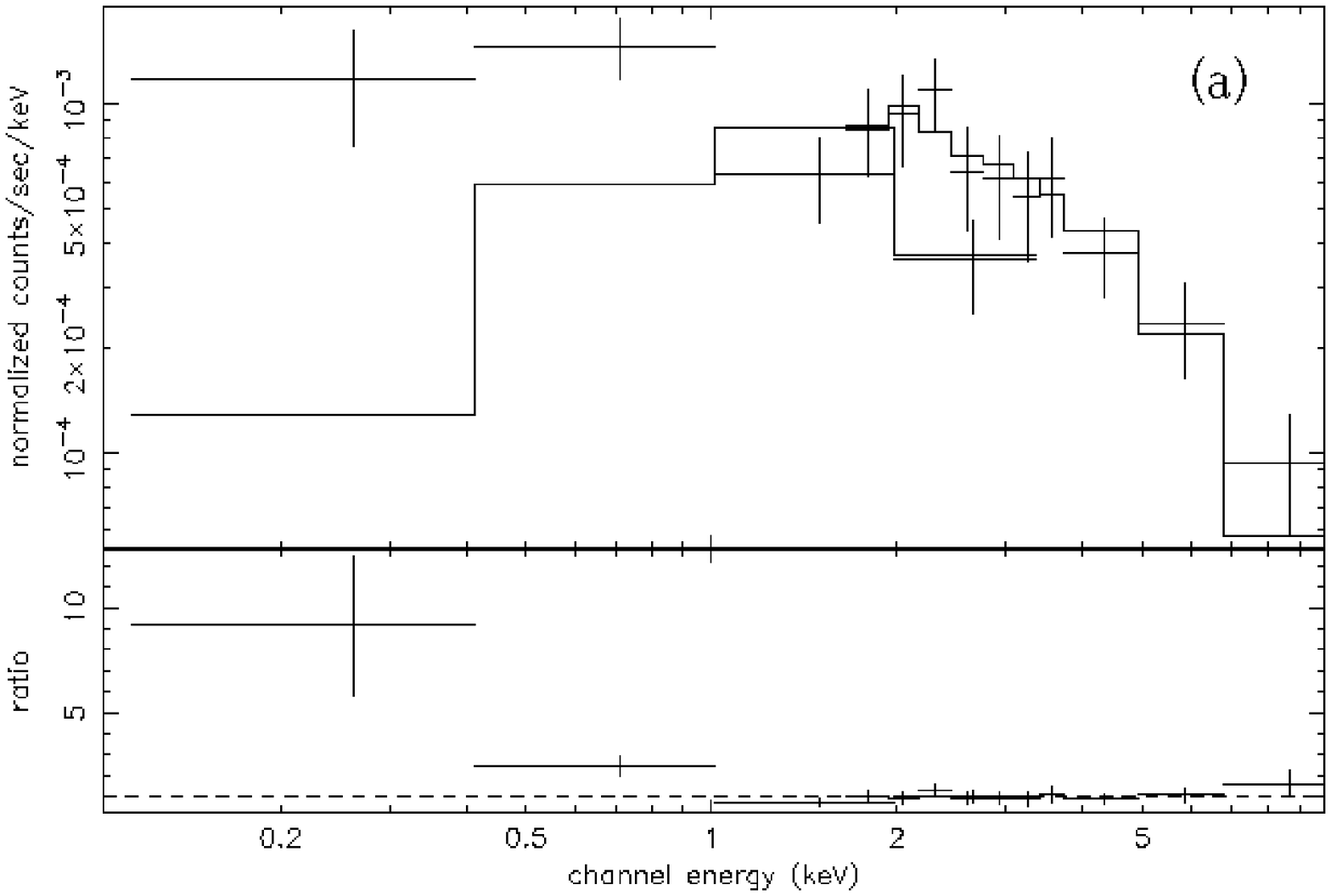} 

\vspace{2cm}

\epsfxsize=9cm 
\hspace{3cm}\epsfbox{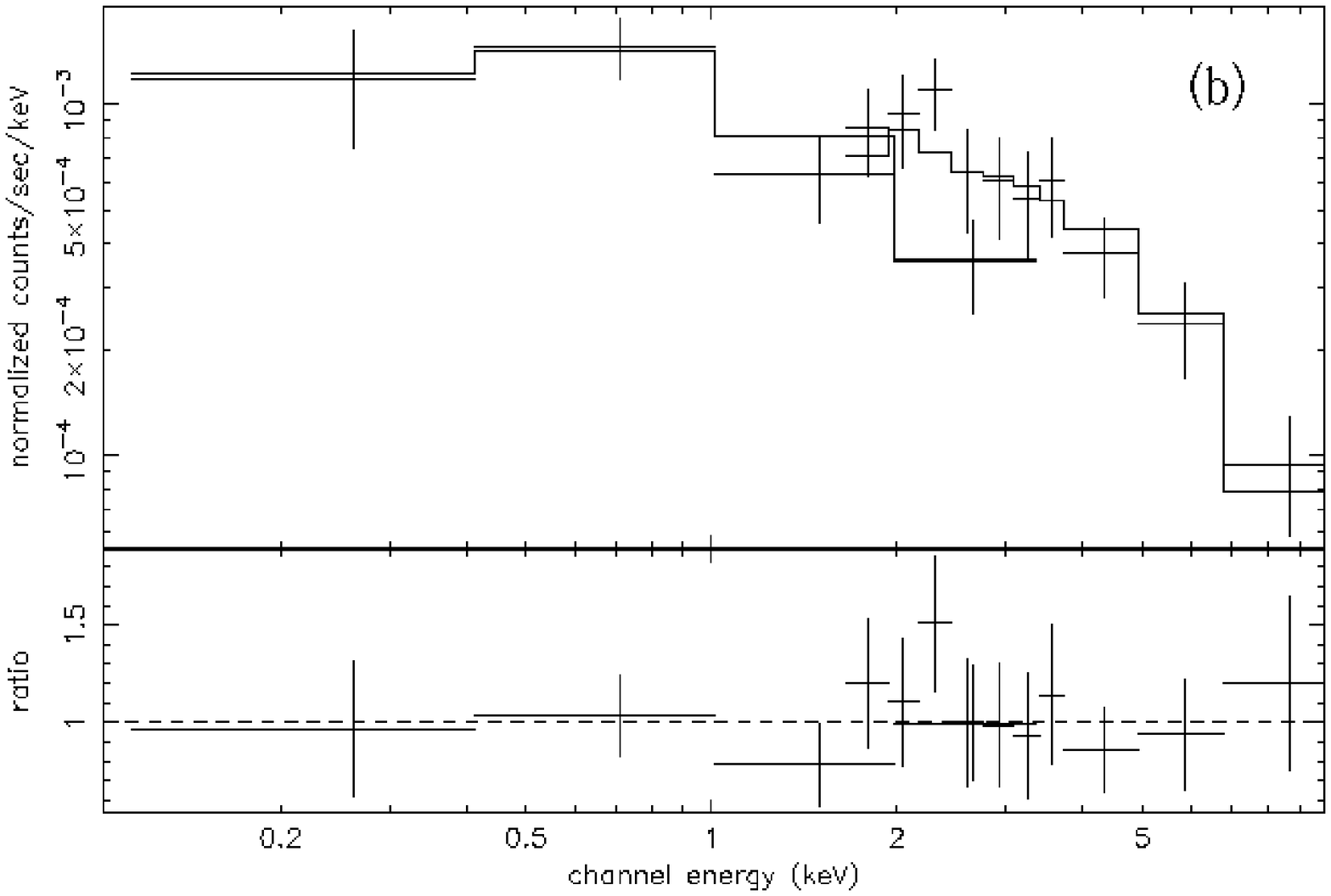} 

\vspace{3cm}

Fig. 5 -- BeppoSAX LECS and MECS spectra of S1 fitted with (a) a thermal
bremsstrahlung and (b) a thermal bremsstrahlung plus black
body.  The lower panels show the ratios
between the data and the model. See Table 2 for the fit parameters.

\end{figure}  

\newpage

\begin{figure}

\vspace{-2cm}

\epsfysize=8cm 
\hspace{3cm}\epsfbox{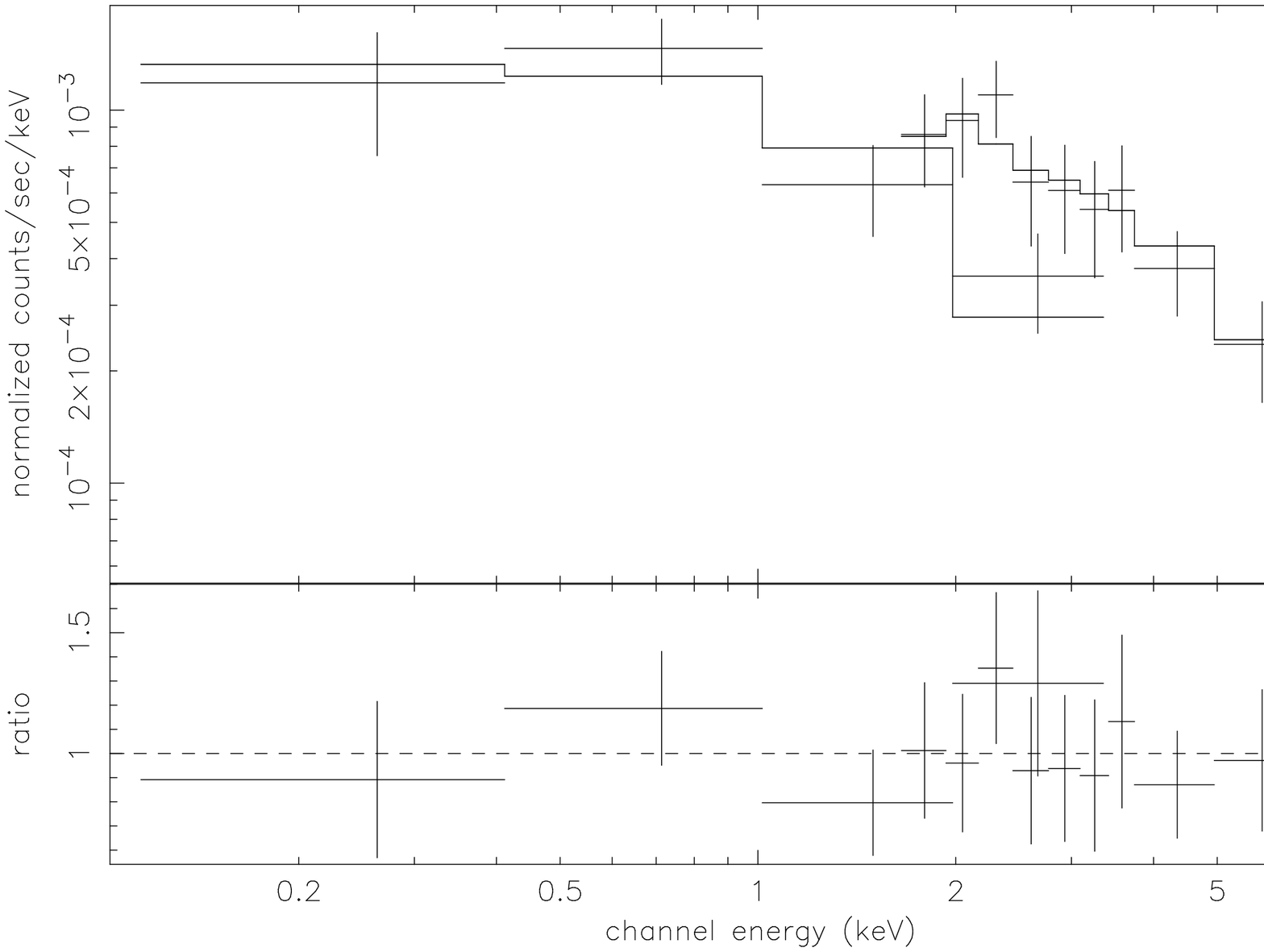} 

\vspace{3cm}

Fig. 6 -- BeppoSAX LECS and MECS spectra of S1 fitted with a broken
power-law. The lower panel shows the ratios between the data and the
model. See Table 2 for the fit parameters.

\end{figure}

\newpage


\newpage

\begin{center}

\begin{figure}
\epsfxsize=9cm 
\hspace{5cm}\epsfbox{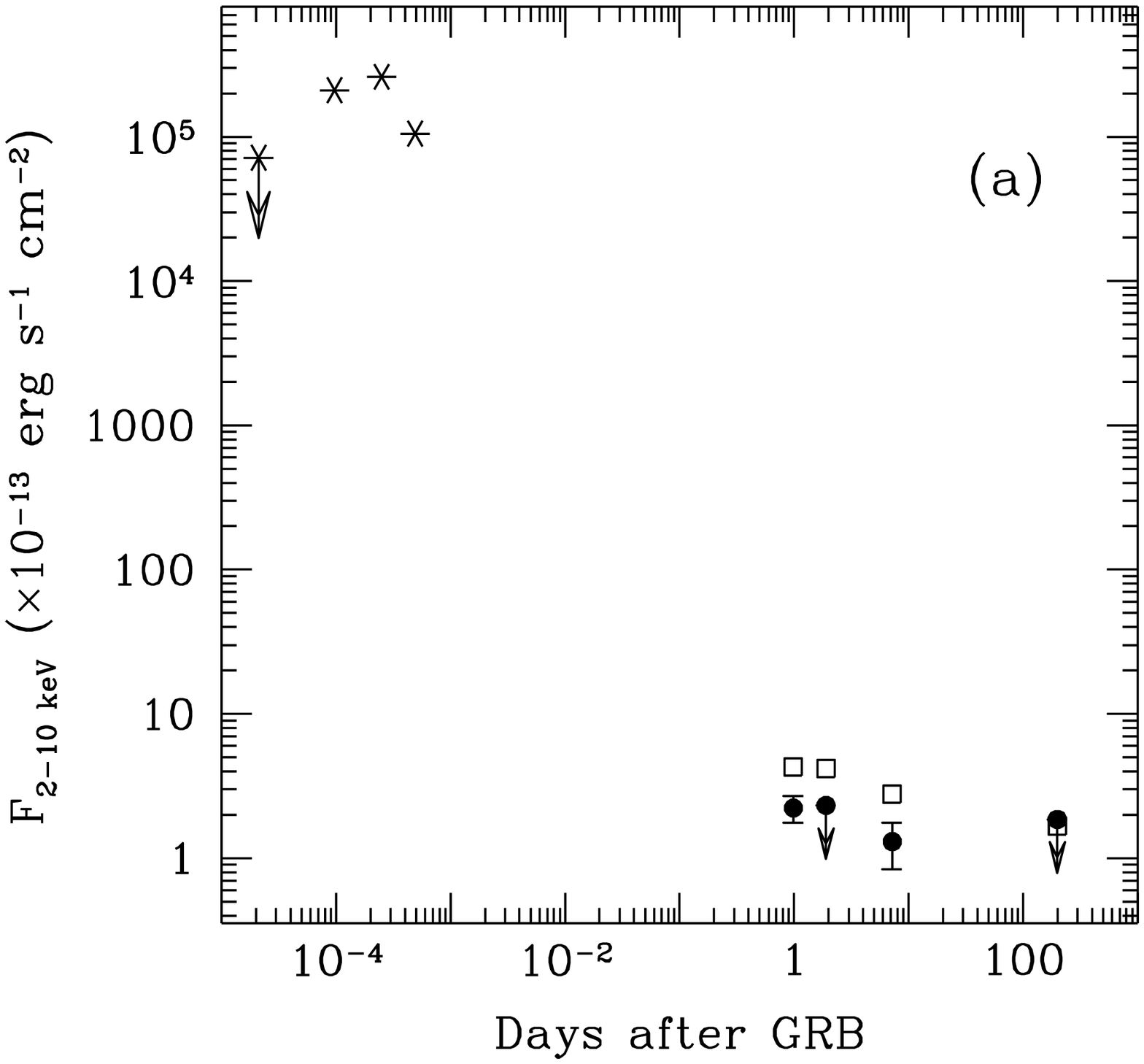} 


\begin{tabular}{cc}
\psfig{figure=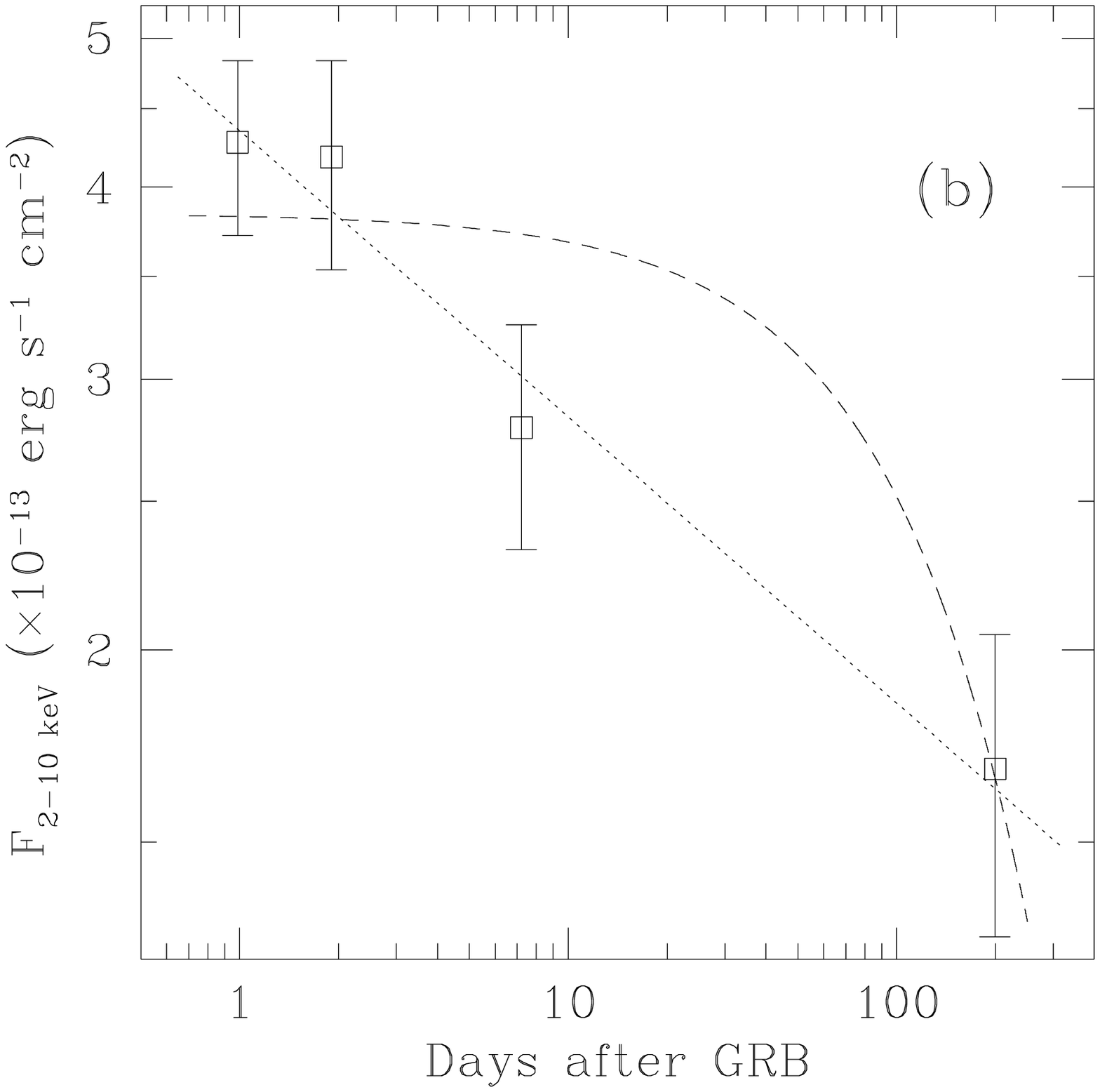,width=8.5cm,height=8.5cm,clip=}
\psfig{figure=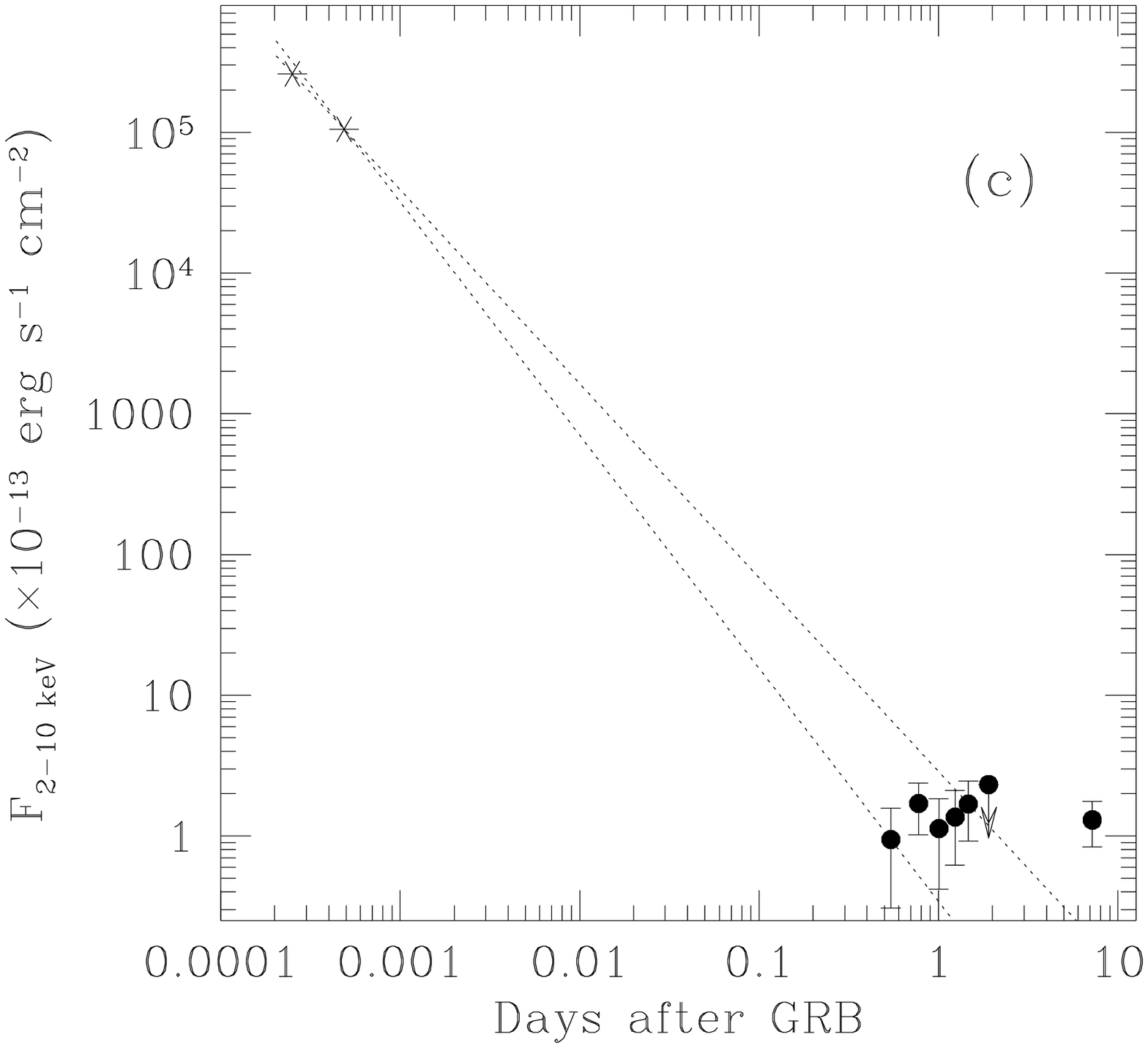,width=8.5cm,height=8.5cm,clip=}
\end{tabular}

Fig. 7 -- BeppoSAX MECS light curves in the 2-10 keV band of the
X-ray sources S1 (open squares) and S2 (filled circles) detected in the
GRB980425 field. The WFC early measurements in the same band are also
shown (stars).  Uncertainties for the NFI measurements are 1-$\sigma$.
The 1-$\sigma$ error bars for the WFC points are smaller than the symbol
size and have not been reported. (b) Same as (a) for source S1 only. The
fits to the temporal decay with a power-law of index $\sim 0.2$ and with
an exponential law of $e$-folding time $\sim 500$ days are shown as
dotted and dashed curves, respectively. (c) Same as (a) for source S2
only. The first NFI measurement of S2 in (a) is replaced here by 
5 points obtained by integrating and averaging the flux in shorter time
intervals. The dotted lines represent the
power-laws of indices $p \simeq 1.6$ and $p \simeq 1.4$ connecting the
last WFC measurement and the first and last of the 5 NFI points of
the April 26-27 light curve, respectively.  The extrapolations of the
power-laws to the time of the third NFI observation (May 1998) fall
below the lower bound of the data point, although they are
compatible with it at the $\simgt 2.5 \sigma$ level.

\end{figure}
\end{center}

\end{document}